\documentclass[bibyear]{aa}  
\usepackage{txfonts}
\usepackage{url}
\usepackage{enumitem}

\usepackage{graphicx}

\def\spose#1{\hbox to 0pt{#1\hss}}
\newcommand\lsim{\mathrel{\spose{\lower 3pt\hbox{$\mathchar"218$}}
     \raise 2.0pt\hbox{$\mathchar"13C$}}}
\newcommand\gsim{\mathrel{\spose{\lower 3pt\hbox{$\mathchar"218$}}
     \raise 2.0pt\hbox{$\mathchar"13E$}}}

\def\ltsima{$\; \buildrel < \over \sim \;$}
\def\lsim{\lower.5ex\hbox{\ltsima}}
\def\gtsima{$\; \buildrel > \over \sim \;$}
\def\gsim{\lower.5ex\hbox{\gtsima}}

\def\apjl{ApJL}
\def\apjs{ApJSS}

\def\aap{A\&A}
\def\sch{Schwarzschild}
\def\ergcms{{\rm\thinspace erg \thinspace cm^{-2} \thinspace s^{-1}}}

\usepackage[normalem]{ulem}

\begin{document}

\title{A NuSTAR view of powerful $\gamma$--ray loud blazars }
\titlerunning{NuSTAR  blazars}
\authorrunning{Ghisellini et al.}
\author{G. Ghisellini\inst{1}\thanks{E--mail: gabriele.ghisellini@brera.inaf.it},
M. Perri\inst{2}$^,$\inst{3}, L. Costamante\inst{2}, G. Tagliaferri\inst{1}, T. Sbarrato\inst{4}, 
S. Campitiello\inst{5}, G. Madejski\inst{6}, F. Tavecchio\inst{1}, G. Ghirlanda\inst{1}  \\
}
\institute{$^1$  INAF -- Osservatorio Astronomico di Brera, Via Bianchi 46, I--23807 Merate, Italy \\
$^2$ Space Science Data Center - Agenzia Spaziale Italiana (SSDC--ASI), via del Politecnico, s.n.c., I-00133, Roma, Italy \\
$^3$ INAF-Osservatorio Astronomico di Roma, Via Frascati 33, I-00078 Monteporzio Catone, Italy \\
$^4$ Dipartimento di Fisica G. Occhialini, Univ. Milano--Bicocca, P.za della Scienza 3, I--20126 Milano, Italy   \\
$^5$ Scuola Internazionale Superiore di Studi Avanzati, Via Bonomea 265, I-34135 Trieste, Italy\\
$^6$ Kavli Institute for Particle Astrophysics and Cosmology, SLAC National Accelerator Laboratory, 
Menlo Park, CA 94025, USA \\
}

\abstract{We observed with the {\it NuSTAR} satellite 3 blazars at $z>2$, detected in the $\gamma$--rays
by {\it Fermi}/LAT and in the soft X--rays, but not yet observed above 10 keV.
The flux and slope of their X--ray continuum, together with {\it Fermi}/LAT data allows us to 
estimate their total electromagnetic output and peak frequency.
For some of them we can study the source in different states, and investigate the main cause
of the observed different spectral energy distribution. 
We then collected all blazars at redshift greater than 2 observed by {\it NuSTAR}, and
confirm that these hard and luminous X--ray blazars are among the most powerful persistent
sources in the Universe.
We confirm the relation between the jet power and the disk luminosity, extending it at
the high energy end.
}
\maketitle

\section{Introduction}

Blazars are radio loud AGNs whose relativistic jet points directly at us, i.e., with a viewing 
angle $\theta_{\rm v}\lsim 1/\Gamma$ with respect to the jet axis, where $\Gamma$ is the jet bulk 
Lorentz factor.  
The jet emission is greatly boosted by relativistic beaming, 
making blazars well visible also at high cosmic distances. 

The beamed non--thermal spectral energy distribution (SED) 
of powerful blazars is characterised by two broad distinctive humps.
Most of the electromagnetic output of very powerful blazars is in the MeV band, 
just where we have no sensitive instrument to look at. 
We can detect them in the adjacent bands, 
through {\it Fermi}/LAT ($>$100 MeV) or in the hard X--rays, 
through {\it INTEGRAL}, {\it Swift}/BAT and {\it NuSTAR}. 
Only {\it NuSTAR} has the spectral resolution (through pointed observations) to accurately find out, 
together with the LAT data (detections and upper limits), the peak frequency 
and luminosity of the blazar emission. 
We claimed (Ghisellini et al. 2010, hereafter G10) that the trend of lower intrinsic peak 
frequency with larger luminosity, observed in blazars of low and intermediate power, 
continues to be valid also at the extreme high power end of the population. 
This was based on blazars detected by BAT, but not by LAT. 
Instead, considering blazars detected by both instruments,
Ajello et al. (2009) claimed that no trend was visible. 
In addition to this controversial intrinsic property, the K-correction
favours in any case the detection in the hard X--ray band of blazars at high redshifts.
Therefore the most powerful persistent objects of the Universe should
be found in the hard X--ray band.
Looking for these extreme objects, we have proposed to observe with {\it NuSTAR}
a few blazars at $z>2$ that have been already detected by {\it Fermi}/LAT, but not by 
{\it Swift}/BAT, hoping to shed light on the intrinsic properties of these sources, 
and in particular on the possible relation between the peak frequency of the high 
energy component of the SED and its luminosity.

Another key question in modern cosmology is how supermassive 
black holes (SMBH) gained most of their mass, especially at the highest 
redshifts probed by current observations. 
Most high--$z$ searches of SMBHs concern radio--quiet objects, but a very promising
alternative approach concerns radio--loud ones, and specifically {\it blazars}.
Beaming makes blazars a unique tool in assessing the number density of radio--loud SMBH at high redshift. 
In fact, for any confirmed high--redshift blazar there must exist other 
$2\Gamma^2 =450 (\Gamma/15)^2$ sources sharing the same intrinsic properties, but whose jets are 
not pointing at us. 
Some SMBHs with masses in excess of $10^9M_\odot$ were already in place 
when the Universe was only $\simeq 700$ Myrs old (e.g., ULAS J1120+0641 at $z=7.08$, 
Mortlock et al. 2011; ULAS J1342+0928  at $z=7.5$, Ba\~nados et al. 2018). 
Their very existence is difficult to reconcile with black hole growth at the 
Eddington rate starting from stellar sized seeds (e.g. Volonteri 2010). 

To the three blazars observed for the first time by {\it NuSTAR},
we have added all other blazars with $z>2$ observed by {\it NuSTAR},
in order to better understand their common properties.
We will show that all of them belong to the group of the most 
powerful blazars both in their jet and in their accretion disk 
properties, fully confirming the fact that the jet power is 
proportional to the accretion luminosity and our expectations
that the hard X--ray selection of high redshift blazars 
picks up the most powerful sources.

We use a flat $\Lambda$CDM cosmology with $h=\Omega_\Lambda=0.7$.

\begin{table*} 
\centering
\footnotesize
\begin{tabular}{lllll lllll ll}
\hline
\hline
RA	 &Dec &Alias &$z$ &$F_{\rm 5}$ &$F_{[0.3-10]}$ &$\Gamma_X$ &$L_X$  &$m_R$ &$M_{\rm BH}^{\rm vir}$   &  \\
     &    &      &    &Jy                &cgs            &     &erg/s  &      &$M_\odot$                &  \\  
\hline   
01 26 42 &$+$25 59 01 &PKS 0123+25   &2.358 &1.4  &2.5e--12 &1.4 &5.2e46 &17.8 &1.8e9   &  &\\ 
02 29 28 &$-$36 43 56 &PKS 0227--369 &2.115 &0.4  &1.3e--12 &1.4 &2.2e46 &19.0 &---     &  &\\ 
05 01 12 &$-$01 59 14 &TXS 0458--020 &2.291 &3.3  &1.4e--12 &1.5 &3.1e46 &19.0 &4.6e8   &  & \\ 
\hline
\hline
\end{tabular}
\caption{\small Selected targets: coordinates (J2000), alias, 
redshift $z$, radio flux at 5 GHz;
X--ray flux in the 0.3--10 keV band;
X--ray photon spectral index; K--corrected 0.3--10 keV luminosity;
$R$ magnitude; 
virial black hole mass.
The virial black hole masses are respectively from: 
Kelly \& Bechtold (2007); Shaw et al.\ (2012); Fan \& Cao (2004); Shen et al.\ (2011).
}
\label{sample}
\end{table*}

\section{Data analysis}

\begin{table*} 
\centering
\centerline{PKS 0123+25}
\vskip 0.1 truecm
\begin{tabular}{llllll}
\hline
\hline
Date &  $\Gamma$ & $F_{\rm 3-5\, kev}$ & $F_{\rm 5-10\, kev}$  & $F_{\rm 10-30\, kev}$  & $\chi^2$ / dof \\
 & & $\ergcms$ & $\ergcms$ & $\ergcms$ & \\
\hline  
\hline
2018 Jan 03 & $1.68_{-0.12}^{+0.12}$ & $4.7\times10^{-13}$ & $7.7\times10^{-13}$ & $1.6\times10^{-12}$ & 13.3 / 30\\
\hline
\hline 
\end{tabular}
\vskip 0.2 truecm
\centerline{PKS 0227--369}
\vskip 0.1 truecm
\begin{tabular}{llllll}
\hline
%
\hline
2017 Aug 10 & $1.35_{-0.26}^{+0.27}$ & $9.6\times10^{-14}$ & $1.9\times10^{-13}$ & $5.5\times10^{-13}$ & 3.5 / 10\\
\hline
\hline 
\end{tabular}
\vskip 0.2 truecm
\centerline{TXS 0458--020}
\vskip 0.1 truecm
\begin{tabular}{llllll}
\hline
%
\hline
2018 Apr 26 & $1.66_{-0.08}^{+0.08}$ & $8.3\times10^{-13}$ & $1.4\times10^{-12}$ & $3.0\times10^{-12}$ & 57.2 / 55\\
\hline
\hline 
\end{tabular}
\vskip 0.2 true cm
\caption{Parameters of the X-ray spectral analysis of the {\it NuSTAR} data. The errors are at 90\% level 
of confidence for one parameter of interest. Fluxes are corrected for the galactic absorption.}
\label{xspec}
\end{table*}

Table \ref{sample} lists the three blazars observed by {\it NuSTAR}, selected among all blazars
at $z>2$ already detected by {\it Fermi}/LAT (Atwood et al. 2009), 
having a [0.3--10 keV] flux larger than $10^{-12}$
erg cm$^{-2}$ s$^{-1}$ and not already observed by {\it NuSTAR}, nor by {\it Swift}/BAT.
This table reports also the redshift, the flux at 5 GHz, the optical magnitude in the $R$ band,
and the estimate of the black hole mass obtained through the virial estimate, when available.

\subsection{NuSTAR}

The {\it NuSTAR} satellite (Harrison et al. 2013) observed PKS 0123+25 on 2018 January 03 
(obsID 60367001002),  PKS 0227--369 on 2017 August 10 (obsID 60367002002) and TXS 0458--020 
on 2018 April 26 (obsID 60367003001).
The total net exposure times were 19.9 ks, 23.3 ks and 20.7 ks, respectively.

The Focal Plane Modules A and B (FPMA and FPMB) data sets were processed with the NuSTARDAS 
software package (v.1.8.0) developed by the ASI Space Science Data Center (SSDC, Italy) in 
collaboration with the California Institute of Technology (Caltech, USA). 
Calibrated and cleaned event files were produced with the {\it nupipeline} task
using the version 20170705 of the {\it NuSTAR} Calibration Database (CALDB).

The three sources were all well detected above the background by the two {\it NuSTAR} hard 
X--ray telescopes up to 30 keV.
The FPMA and FPMB energy spectra of the three sources were extracted from the cleaned and
calibrated event files using a circular spatial region with a radius of 12 pixels 
($\sim 30$ arcseconds) centered on the target, while the background was extracted from nearby
circular regions of 50 pixel radius. 
The ancillary response files were generated with the {\it nuproducts} task, applying corrections 
for the Point Spread Function (PSF) losses, exposure maps and telescope vignetting. 

For all three observations the spectral analysis of the {\it NuSTAR} data was performed using 
the XSPEC package adopting a single power-law model with an absorption hydrogen--equivalent column 
density fixed to the Galactic values given by Kalberla et al. (2005), 
i.e. $N_{\rm H} = 6.8\times 10^{20}$~cm$^{-2}$ for PKS 0123+25 , $N_{\rm H} = 2.4\times 10^{20}$~cm$^{-2}$
for PKS 0227--369 and $N_{\rm H} = 6.0\times 10^{20}$~cm$^{-2}$ for TXS 0458--02. 
All spectra were binned to ensure a minimum of 30 counts per bin and energy channels below 3.0 keV 
and above 30.0 keV were excluded. 
A multiplicative constant factor was included to take into account for cross-calibration uncertainties 
between the two telescopes ({\it NuSTAR} FPMA and FPMB). 
We found that this model fit the spectral data very well for all three sources in the considered energy band. 
The results of the spectral fits are given in Table \ref{xspec}.

\subsection{Swift-XRT}

The {\it Neil Gehrels Swift Observatory} (Gehrels et al. 2004) observed with the X-ray Telescope 
(XRT, Burrows et al. 2005) the source PKS 0123+25 simultaneously with {\it NuSTAR}, 
namely on 2018 January 03 and January 4 (obsIDs 00088100001, 00088100002), for a total net exposure 
time of 2.0 ks.

The XRT observations were carried out with the Photon Counting (PC) readout mode. 
The XRT data were first processed using the XRT Data Analysis Software (XRTDAS, v.3.4.1), 
which was developed under the responsibility of the ASI Space Science Data Center. 
Standard calibration and cleaning processing steps were applied using the {\it xrtpipeline} 
software module and using the version 20180710 of the {\it Swift-XRT} Calibration Database (CALDB).

Source events for the spectral analysis were extracted in the 0.3--10 keV energy band using 
a circular spatial extraction region with a 20 pixels radius ($\sim 47$ arcseconds). 
The background was estimated using a nearby source-free circular region with a radius of 50 pixels.
Corrections to the ancillary response files for PSF losses, CCD defects and telescope 
vignetting were calculated and applied using the {\it xrtmkarf} software module.

For the spectral analysis the energy spectrum was grouped to ensure at least 20 counts in each bin. 
We adopted an emission model described by a single power-law with an absorption hydrogen-equivalent 
column density fixed to the Galactic value of $N_{\rm H} = 6.8\times 10^{20}$~cm$^{-2}$
(Kalberla et al. 2005). 
The results of the spectral fit were found to be consistent in slope and normalisation with the ones 
derived from the {\it NuSTAR} observation, thus extending the observed spectral slope down 
to 0.3 keV, with a best fit  photon index $\Gamma = 1.7_{-0.3}^{+0.3}$.

For the two blazars PKS 0227--369 and TXS 0458--020 no simultaneous observations with {\it NuSTAR} 
were carried out by the {\it Neil Gehrels Swift Observatory}.

\subsection{Fermi/LAT}

We analyzed the {\it Fermi}/LAT data around the {\it NuSTAR} pointings 
using the Pass--8 data version  and the public Fermi Science Tools version v11r5p3.

First we looked for nearly simultaneous data, with several choices of
exposure time, until we derived a detection. 
The blazar TXS 0458--02 was in a bright state, and an integration time of just 2 days 
($\pm$1 day around the {\it NuSTAR} pointing) was enough for a detection of $\sim11\sigma$. 
The other two objects, instead, require years of integration for a detection. 
We therefore considered two exposures, a short one of 30 days 
($\pm$15 days around the {\it NuSTAR} pointing) to derive a meaningful upper limit at 
the same epoch of {\it NuSTAR}, and a long one of years, in order to measure the average spectrum. 
The long exposure is 4 years for PKS 0123+25 (from May 24, 2014 to May 24, 2018),   and 
2 years for PKS 0227--369 (from May 24, 2016 to May 24).
The results are reported in Table \ref{lat}.

Gamma--ray events were selected from a Region of Interest (ROI) of 15\degr using
standard quality criteria, as recommended by the Fermi Science Support Center (FSSC).
We performed the likelihood analysis in two steps.
In the first step the XML model included all the sources in the 
preliminary LAT 8--year Point Source List (FL8Y). 
We then performed a second  likelihood fit using the XML model from the first step, 
optimized by dropping all sources with a TS$<1$.
The analysis was performed with the NEWMINUIT optimizer, using an unbinned likelihood 
for the short datasets and a binned likelihood for the long exposures, with 0.1\degr bins 
and 10 bins for decade in energy.

The LAT data points for the SED were obtained by binning the spectrum with 2 bins per decade in 
energy, in the 0.1-100 GeV range, and performing a likelihood analysis in each single energy bin.
In the XML model all parameters were kept fixed to the best--fit values, except 
for the normalization of the target and of the two backgrounds (isotropic and galactic).  
A binned or unbinned likelihood was used if the total number of counts in the bin was higher 
or lower than 15000, respectively.
A Bayesian upper limit was calculated if in that bin the target had a TS$<9$ or {\it npred}$<3$.
The light curves were obtained by performing an unbinned likelihood analysis in each time bin of 
7 days, leaving free the parameters of the brightest or variable FL8Y sources in the ROI, 
within an 8--degree radius of the target.  

\begin{table}
\centering
\label{lat}
\begin{tabular}{lrccc}  
\hline
\hline
Source         &   TS      &  Flux           &  $\Gamma_{\rm LAT}$  & Exp.    \\
~[1]           &   [2]     &    [3]          &       [4]            & [5]         \\%
\hline 
PKS 0123+25    &     0.0   &  $< 2.18$       &  2.8            & 30d \\
               &   32.7    &  $1.28\pm0.35$  &  $2.89\pm0.23$  & 4y  \\
PKS 0227--369  &     0.0   &  $< 1.38$       &  2.7            & 30d \\
               &    63.3   &  $1.61\pm0.29$  &  $2.73\pm0.14$  & 2y  \\
TXS 0458--02   &   137.8   &  $46.2\pm8.2$\enspace   &  $2.30\pm0.14$  & 2d  \\
\hline		 
\hline  	 
\end{tabular}
\vskip 0.2 true cm	 
\caption{Parameters of the power-law fits to the {\it Fermi}/LAT data.
Col. [1]: object name.
Col. [2]: test statistics (Mattox et al. 1996).
Col. [3]: integrated photon flux or 95\% upper limit in the 0.1-300 GeV band, in units of $10^{-8}$ cm$^{-2}$ s$^{-1}$. 
Col. [4]: photon index of the LAT spectrum, measured or assumed for the upper limit.
Col. [5]: total LAT exposure, around the {\it NuSTAR} pointing, in days (d) or years (y).
}
\end{table}

\section{Modelling}
\label{model}

We interpret the overall SEDs of our sources with a leptonic, one--zone 
jet emission model plus the contribution from an accretion disk, its X--ray corona, 
and a molecular torus, that is absorbing and re--emitting in 
the infrared a fraction of the disk radiation.
The detail of the model are in Ghisellini \& Tavecchio (2009) and Ghisellini and Tavecchio (2015)
and we summarize here its main features.

\begin{itemize}

\item The emitting region producing the non--thermal radiation is assumed
to be spherical, with radius $R$ and at a distance $R_{\rm diss}$ from the central black hole.
The jet is assumed conical, with semi--aperture angle $\psi$.
Although $\psi\Gamma\sim 1$ is born out by numerical simulations of jet acceleration, 
jets could have a parabolic shape while accelerating, becoming conical when coasting
(e.g. Marsher 1980; Komissarov et al. 2007). 
They could also re--collimate at large distances, making the relation between 
the transverse radius $r$ and the distance $R_{\rm diss}$ uncertain.
We assume, for simplicity, $\psi=0.1$, corresponding to $5.7^\circ$
and, roughly, $\psi\sim 1/\Gamma$.
The emitting plasma is assumed to move with a bulk motion of velocity $\beta c$ and Lorentz factor
$\Gamma$ at a viewing angle $\theta_{\rm v}$ from the line of sight.
The Doppler
factor is $\delta = 1/[\Gamma(1-\beta\cos\theta_{\rm v})]$.

\item Throughout the emitting region relativistic electron are continuously 
injected at a rate
$Q(\gamma)$ [cm$^{-3}$ s$^{-1}$] for a time equal to the 
light crossing time $R/c$. 
The shape of $Q(\gamma)$ is assumed to be a smoothly broken power law,
with a break at $\gamma_{\rm b}$:
\begin{equation}
Q(\gamma)  \, = \, Q_0\, { (\gamma/\gamma_{\rm b})^{-s_1} \over 1+
(\gamma/\gamma_{\rm b})^{-s_1+s_2} } \,\, {\rm cm^{-3}}
\label{qgamma}
\end{equation}
\item
The power injected in the form of relativistic electrons is
\begin{equation}
P_{\rm inj}^\prime  \, =\, m_{\rm e}c^2 \int Q(\gamma)\gamma d\gamma
\label{pprime}
\end{equation}
This is calculated in the comoving frame.
We solve the continuity equation to find the energy distribution $N(\gamma)$ [cm$^{-3}$]
of the emitting particles at the particular time $R/c$, when the injection process 
is assumed to end. 
We account for synchrotron and inverse Compton cooling and $e^\pm$ pair production and reprocessing,
although, in our sources, $e^\pm$ pairs are not important.

\item
The magnetic field $B$ is tangled and uniform throughout the emitting region.
\item
There are several sources of radiation externally to the jet:
\begin{enumerate}
\item the broad line photons, assumed to re--emit 10\% of the accretion luminosity
from a shell--like distribution of clouds located at a distance 
$R_{\rm BLR}=10^{17}L_{\rm d, 45}^{1/2}$ cm;
\item the IR emission from a dusty torus, located at a distance
$R_{\rm IR}=2.5\times 10^{18}L_{\rm d, 45}^{1/2}$ cm;
\item  the direct emission from the accretion disk, including its X--ray corona;
\item  the starlight contribution from the inner region of the host galaxy and 
the cosmic background radiation.
\end{enumerate}

All these contributions are evaluated in the blob comoving frame, where we calculate the 
corresponding inverse Compton radiation from all these contributions, and then transform
into the observer frame.

\item The numerical code we use is not time dependent: it gives 
a ``snapshot" of the 
predicted SED at the time $R/c$, when the particle distribution $N(\gamma)$ 
and consequently the produced flux are at their maximum.

\item
For powerful sources, the radiative cooling is efficient, and the cooling
timescale can be shorter than $R/c$ even for the low energy particles.
This implies that $\gamma_{\rm peak}$, the random Lorentz factor of the electron emitting
most of the radiation is close to $\gamma_{\rm b}$.


\item
The size of the emitting region is rather compact, as indicated by the short variability 
timescales observed in blazars.
As a consequence, the synchrotron flux is self--absorbed at high frequencies,
in the submm band.
Therefore the model cannot account for the radio emission at lower frequencies,
that must be produced by more extended regions of the jet.

\item
To calculate the flux produced by the accretion disk, we adopt a standard 
Shakura \& Sunyaev (1973) disk (see Ghisellini \& Tavecchio 2009).
This model depends mainly on the accretion rate (regulating the total disk luminosity)
and on black hole mass (regulating the location of the peak of the emission).
This allows us to fit also the thermal radiation seen in the optical--UV range, and
to estimate the accretion rate and the black hole mass.

\item The disk luminosity is independent of the adopted accretion model (e.g.
standard Shakura \& Sunyaev, with zero spin, or an accretion disk around a Kerr
black hole).
Instead the estimate of the mass does depend on the assumed accretion model 
(see e.g. Calderone et al. 2013. See also Campitiello et al. (2018) that
studied how the black hole spin and the special and general relativistic effects impact 
on the determination of the black hole mass).
          
\item The total jet power is the sum of the power carried by particles (we assumed one cold proton
per emitting electron), magnetic field and radiation.
Therefore the estimate of the magnetic and particle power is model dependent, because 
the particle number and the value of the magnetic field depend on which model we are
using to interpret the data (leptonic or hadronic, molti or one--zone, and so on).
This is calculated at the dissipation region, through
\begin{equation}
P_{\rm i} = \pi \psi^2 R^2_{\rm diss} U_{\rm i}\Gamma^2\beta c
\end{equation}
where the subscript ``i" can stand for protons, electrons, magnetic field, or radiation,
and $U$ is the corresponding energy density, as calculated in the comoving frame.
The power in radiation is instead model independent. 
It can be calculated with the equation above, that can be re-written as 
(for viewing angles $\theta_{\rm v} \sim 1/\Gamma$):
\begin{equation}
P_{\rm r} \, \sim \, 2\, {L^{\rm bol}_{\rm jet} \over \Gamma^2}
\end{equation}
where $L^{\rm bol}_{\rm jet}$ is the bolometric observed luminosity produced by the jet.
This is an observable. 
Therefore only the knowledge of $\Gamma$ enters this estimate. 
This makes $P_{\rm r}$ almost model independent. 
It is a {\it lower limit} of the jet power.
$P_{\rm jet}$ is the sum of the different components.

\item
The uniqueness of the parameter values has been discussed in some detail in Ghisellini \& Tavecchio (2015).
We have stressed there that in the framework of our leptonic, one--zone model, it is possible to find a 
unique solution for fitting the SED, but only if the data are of sufficient quality. 
One would need simultaneous data from the mm to the $\gamma$--rays, and this is possible only in a few cases.
We are then constrained to assume that the not--simultaneous data we have collected are a reasonably
good representation of the SED. 
We have tried to constrain the $\gamma$--ray flux and slope the best we could, by analyzing the 
{\it Fermi}/LAT data as close as possible to the{\it NuSTAR} observations.
In addition, when possible, we will compare the resulting SED with the SED corresponding to other states of the sources,
to enlighten the possible causes of variations.

\end{itemize}

\section{Results}

\begin{figure} 
\vskip -0.6 cm
\includegraphics[width=9cm]{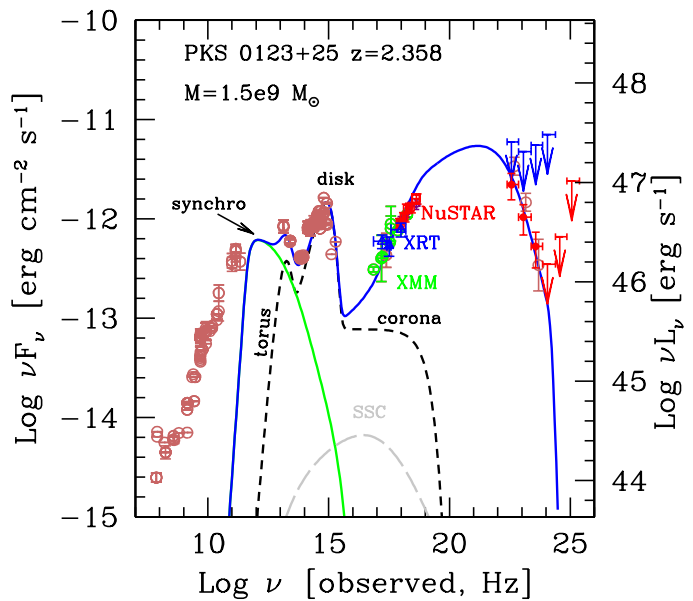} 
\vskip -0.6 cm
\caption{The overall SEDs of PKS 0123+25.
Besides our data (red points), 
we show the archival data collected from the ASI/SSDC database.
We have indicated in green the XMM--{\it Newton} data,
taken in January 2009, and in blue the {\it Swift}/XRT data
taken simultaneously with the {\it NuSTAR} observation.
The blue arrows in the $\gamma$--ray band are upper limits
obtained integrating over 30 days (15 days before and 15 days after 
the {\it NuSTAR} observation).
Red $\gamma$--ray points and arrows correspond to the average flux
during the last 4 years.
The lines are the result of the modelling (see text).
} 
\label{sed0126}
\end{figure}

We show the overall SEDs of the three blazars analyzed in this paper in Figs.
\ref{sed0126}, \ref{sed0227} and \ref{sed0458}.
The SEDs of PKS 0123+25 and PKS 0227--369 show the presence of
a thermal component at optical--UV frequencies, that we interpret as the
due to a standard accretion disk.
Perhaps more surprising, this thermal emission is not clearly visible 
in TXS 0458--02, most probably because it is hidden by the dominating
synchrotron spectrum.
Besides showing our data, the figures reports the archival data from 
the ASI/SSDC database (https://tools.ssdc.asi.it/).

\begin{figure} 
\vskip -0.6 cm
\includegraphics[width=9cm]{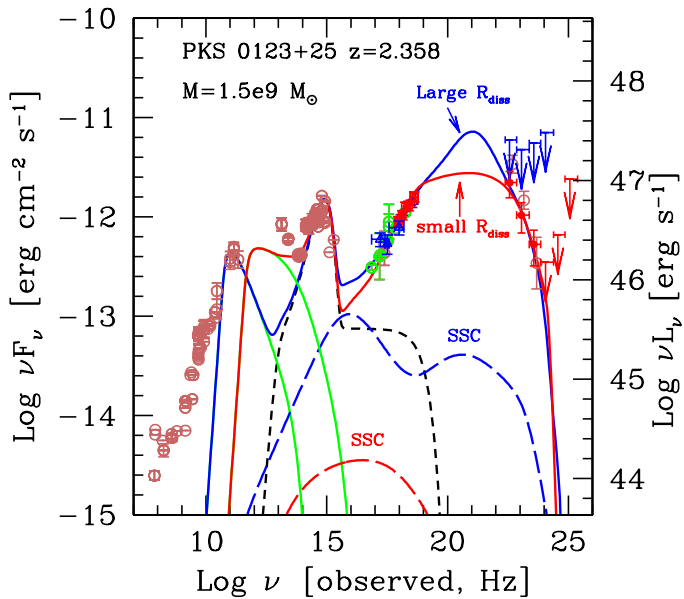} 
\vskip -0.6 cm
\caption{
The lines are the result of the modelling assuming that there is no torus,
assuming both a small $R_{\rm diss}$ ($=2.25\times 10^{17}$ cm) and a large $R_{\rm diss}$ ($=3.6\times 10^{18}$ cm). 
Parameters are listed in Table \ref{para} and Table \ref{para2}.
If we fit the high energy emission, the model underproduces the near IR flux.
Note that even when $R_{\rm diss}$ is large, the high energy emission is dominated
by the external Compton, while the SSC flux (labelled, the first and second order Compton are shown) is much weaker.
This is because the BLR emission, even if severely un--beamed, still dominates
over the internally produced comoving synchrotron energy density.
} 
\label{sed0126notorus}
\end{figure}
\begin{figure} 
\vskip -0.6 cm
\includegraphics[width=9cm]{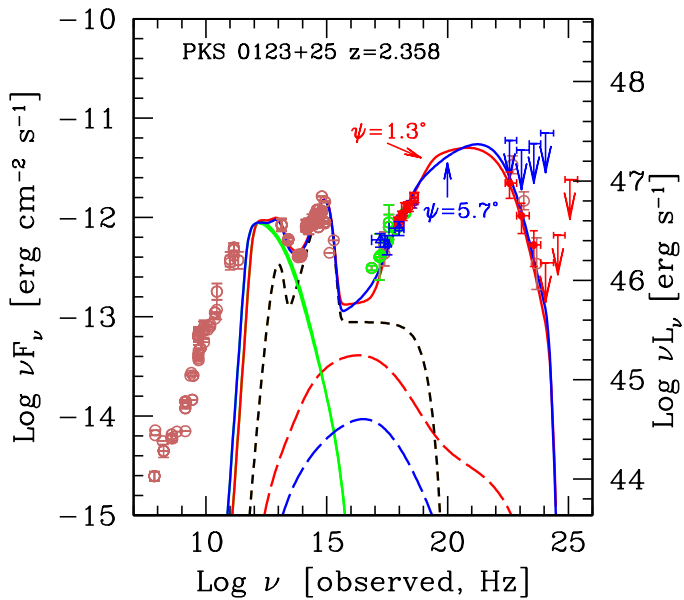} 
\vskip -0.6 cm
\caption{
Comparison between the best models assuming $\psi=0.1=5.7^\circ$ and $\psi=0.023=1.3^\circ$,
as labelled. The long--dashed lines are the SSC contribution.
Parameters are listed in Table \ref{para} and Table \ref{para2}.
} 
\label{sed0126psi}
\end{figure}

\subsection{PKS 0123+25}

The {\it NuSTAR} data of this source lie on the extrapolation
of the lower energy X--ray data taken by XMM/{\it Newton}  
on Jan. 8, 2009, and the {\it Swift}/XRT data taken simultaneously
with {\it NuSTAR}.
Integrating the {\it Fermi}/LAT data 15 days before plus 15 day after 
the {\it NuSTAR} observation, the source was not detected.
The corresponding 95\%
upper limits are shown in Fig. \ref{sed0126} together with
the {\it Fermi}/LAT spectrum integrating over the last 4 years.
The upper limits are consistent with the spectrum obtained with the long exposure, 
indicating no flares during the {\it NuSTAR} observation.

The optical spectrum can be well fitted by a standard accretion disk model, 
and we find a black hole mass of $M=1.5\times 10^9 M_\odot$ and a 
disk luminosity $L_{\rm d} = 5.85\times 10^{46}$ erg s$^{-1}$,
corresponding to 30\% of the Eddington luminosity.
This value agrees with the observed broad line luminosities,
as observed by the SDSS spectrum (DR13).
We have used the template of Francis et al. (1991), and 
assumed that $L_{\rm d}=10L_{\rm BLR}$.  
In this way we derived 
$L_{\rm BLR}=10^{46}$ erg s$^{-1}$ (using the CIV line);
$L_{\rm BLR}=7.3\times 10^{45}$ erg s$^{-1}$ (CIII] line) and
$L_{\rm BLR}=1.6\times 10^{45}$ erg s$^{-1}$ (MgII line).
In the infrared band there can be the contribution of both the
torus and the jet emission.
In order to disentangle the two, we have assumed that the time averaged 
$\gamma$--ray spectrum is indicative of the high energy emission during the
{\it NuSTAR} observation.
In Fig. \ref{sed0126notorus} we show the model SED assuming there is no torus:
if we fit the high energy SED, we under--reproduce the near IR.
We therefore assume that the near IR flux is produced by the torus, 
and this helps to find the peak of the high energy SED and its dominance with respect
to the synchrotron component. 
These information help to constrain the magnetic field and $\gamma_{\rm peak}$
allowing to find a robust solution for the model parameters
(assuming that the archival data are indicative of the real SED).
Fig. \ref{sed0126psi} compares the models assuming two different values 
for the aperture angle of the jet: $\psi=0.1=5.7^\circ$ (blue lines)
and $\psi=0.023=1.3^\circ$ (red lines).
The latter value corresponds to the average value of {\it Fermi}/LAT blazars
derived by Pushkarev et al. (2017).
Bot models represent the data well, and are indistinguishable.
The model with the smaller $\psi$ requires a larger $R_{\rm diss}$
(factor 3) and a larger jet power (factor 3).
For homogeneity with the blazars fitted previously, in the rest of the paper we
use $\psi=0.1$.   
The parameters are listed in Table \ref{para} and Table \ref{para2}.

\begin{figure} 
\vskip -0.6 cm
\includegraphics[width=9cm]{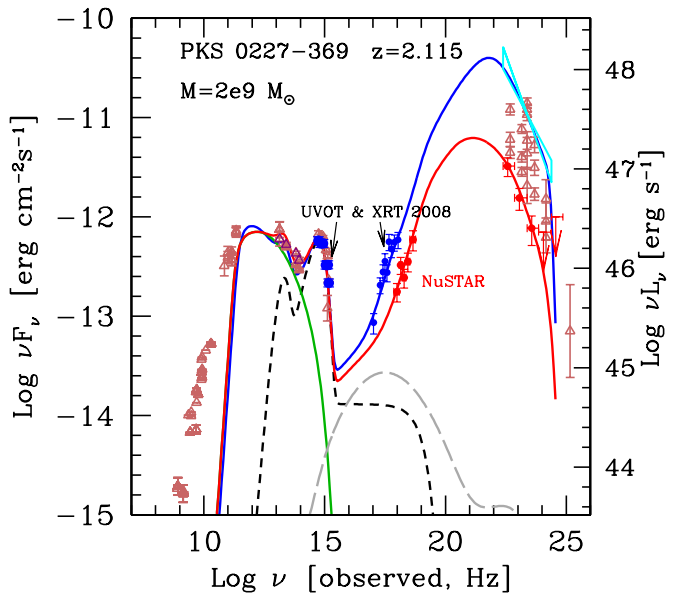}  
\vskip -0.6 cm
\caption{The overall SEDs of PKS 0227--369.
The X-ray flux was significantly lower during the {\it NuSTAR}
observations with respect to an earlier {\it Swift}/XRT+UVOT  observation carried out in November 2008.
The red points in the {\it Fermi}/LAT band correspond to integrating
the last two years of observations.
This shows that the source was in a low state during this period of time.
} 
\label{sed0227}
\end{figure}

\begin{table*} 
\centering
\begin{tabular}{lllll lllll lllll ll}
\hline
\hline
Source  &$z$ &$M^{\rm SED}_{\rm BH}$ &$L_{\rm d}$ &$L_{\rm T}/L_{\rm d}$ &$R_{\rm diss}$ &$R_{\rm BLR}$ &$R_{\rm T}$&$P^\prime_{\rm inj}$   \\
        & &$M_\odot$  &$10^{45}$erg s$^{-1}$ & &$10^{15}$ cm &$10^{15}$ cm &$10^{15}$ cm &$10^{45}$ erg s$^{-1}$ 
         \\ 
\hline
PKS 0123+25 	    &2.358 &1.5e9  &58.5 &0.3 &270 &764 &3.1e4 &0.017     \\        
PKS 0123+25 no torus &2.358 &1.5e9 &58.5 &0   &225 &764 &---   &0.012    \\       
PKS 0123+25 no torus &2.358 &1.5e9 &58.5 &0   &3.6e3 &764 &--- &0.3      \\      
PKS 0123+25 ($\psi=1.3^\circ)$ 	    
                     &2.358 &1.5e9  &58.5 &0.3 &540 &764 &3.1e4 &0.025   \\   
PKS 0227--369 new   &2.115 &2e9    &18.2 &0.5 &660 &427 &5.6e3 &0.011  \\   
PKS 0227--369 old   &2.115 &2e9    &18.2 &0.5 &480 &427 &5.6e3 &0.045    \\  
TXS 0458--020 new   &2.291 &8e8    &10.4 &0.5 &144 &322 &2.7e3 &0.11     \\   
TXS 0458--020 quiesc.  &2.291 &8e8 &10.4 &0.5 &132 &322 &2.7e3 &0.025    \\   
TXS 0458--020 ``flare" &2.291 &8e8 &10.4 &0.5 &192 &322 &2.7e3 &0.35    \\  
TXS 0458--020~~$(\Gamma=7)$ &2.291 &8e8 &10.4 &0.5 &192 &322 &2.7e3 &0.25    \\  
\hline
\hline 
\end{tabular}
\vskip 0.2 true cm
\caption{
Parameters for the models shown in 
Fig. \ref{sed0126}, Fig. \ref{sed0126notorus}, Fig. \ref{sed0126psi}, Fig. \ref{sed0227} and Fig. \ref{sed0458}.
For the BLR we have always assumed  $L_{\rm BLR}=0.1 L_{\rm d}$.  
For all models we assumed$\psi=0.1=5.7^\circ$, except otherwise noted.
For a simple geometry (a spherical torus surrounding the disk), the ratio $L_{\rm T}/L_{\rm d}$ corresponds 
to the aperture angle $\theta_{\rm T}$ of the torus (the angle between the normal to the disk and the border of the torus): 
$L_{\rm T}/L_{\rm d}=\cos^2\theta_{\rm T}$.
A ratio $L_{\rm T}/L_{\rm d}=0.3$ gives $\theta_{\rm T}=57^\circ$, while $L_{\rm T}/L_{\rm d}=0.5$ gives $\theta_{\rm T}=45^\circ$.
}
\label{para}
\end{table*}
\begin{table*} 
\centering
\begin{tabular}{lllll lllll lllll ll}
\hline
\hline
Source   &$B$ &$\Gamma$ 
&$\theta_{\rm v}$ &$\gamma_{\rm b}$ &$\gamma_{\rm max}$ &$s_1$ &$s_2$ &$\gamma_{\rm peak}$ &$\log P_{\rm r}$ &$\log P_{\rm jet}$ & &  \\
         &G& &deg  \\ 
\hline
PKS 0123+25          &6.0 &11 &3 &400 &5e3 &1.5 &4   &98    &45.6 &47.2\\      
PKS 0123+25 no torus &6.6 &12 &3 &1e3 &5e3 &1.9 &4.4 &73    &45.5 &47.4 \\     
PKS 0123+25 no torus &0.036 &22 &2 &200 &5e3 &1.9 &4.4 &181 &45.5 &49.5\\    
PKS 0123+25 ($\psi=1.3^\circ)$ 	
                     &6.7 &11 &3 &550 &5e3 &1.9 &4.3 &54    &45.6 &47.7 \\                 
PKS 0227--369 new   &0.9 &13 &3 &600 &5e3 &1   &3.1 &305    &45.6 &46.5 \\   
PKS 0227--369 old   &1.3 &13 &3 &250 &5e3 &0   &3   &181    &46.3 &46.9 \\   
TXS 0458--020 new   &3.2 &14 &3 &300 &4e3 &--1 &2.5 &317    &46.8 &47.3 \\   
TXS 0458--020 quiesc.  &8.1 &13 &3 &190 &3e3 &0.7 &3 &116   &46.0 &47.1 \\  
TXS 0458--020 ``flare" &2.5 &18 &3 &200 &4e3 &--1 &3 &170   &47.5 &48.2 \\ 
TXS 0458--020$~~(\Gamma=7)$ &1.7 &7 &3 &800 &7e3 &--1 &2.5 &824  &46.5 &47.0 \\ 
\hline
\hline 
\end{tabular}
\vskip 0.2 true cm
\caption{
{\it continue}. Parameters for the models shown in 
Fig. \ref{sed0126}, Fig. \ref{sed0126notorus}, Fig. \ref{sed0126psi}, Fig. \ref{sed0227} and Fig. \ref{sed0458}.
Luminosities are in units of erg s$^{-1}$.
}
\label{para2}
\end{table*}

\subsection{PKS 0227--369}

The X--ray flux was significantly lower during the {\it NuSTAR} observations with respect to an earlier 
{\it Swift}/XRT observation carried out in November 2008 (Ghisellini, Tavecchio \& Ghirlanda 2009).
The shown $\gamma$--ray data (red symbols) refer to the last 2 years, 
and indicate a low state both with respect to the archival data and to
an older flaring state. 
The slopes of both the X--ray and the $\gamma$--ray data are instead the same
as the ones derived by the archival data.
Unfortunately, during the {\it NuSTAR} observations, the source was not observed by
{\it Swift}, so that we cannot check if any change occurred also in the optical--UV bands.
However, we do not expect any strong flux variability in these bands, since they are
produced by the accretion disk, whose emission is usually much more stable than the jet one. 
Applying our standard disk model we derive $M=2\times 10^9 M_\odot$ and 
$L_{\rm d}=1.8 \times 10^{46}$ erg s$^{-1}$, corresponding to 7\% of the Eddington
luminosity.
We did not find any published optical spectra reporting the luminosity of the broad lines.
However, the disk emission is clearly visible in this source and the accretion disk luminosity
we found is therefore reliable.
As in PKS 0123+25, the infrared flux is dominated by the jet synchrotron emission.
As a consequence, the torus component is somewhat uncertain: in Fig. \ref{sed0227}
we show a torus reprocessing half of the disk luminosity.

To model the source, we have assumed that the radio--to--optical archival data 
give a good representation of the SED in this frequency range, and we tried
to explain the change of the SED by changing the minimum number of parameters.

We find that the observed variability can be explained by changing the power of the relativistic 
electrons injected throughout the source, that are responsible for the emission.
The shown models differ by a factor 4 in $P^\prime_{\rm inj}$. 
Furthermore, the lower {\it NuSTAR} state is characterized 
by a slightly larger dissipation region, with a slightly smaller magnetic field and a larger 
value of the energy of the electrons emitting at the peaks of the SED.
The total jet power is a factor three smaller than in the high state.

\begin{figure} 
 \vskip -0.3 cm
\includegraphics[width=9cm]{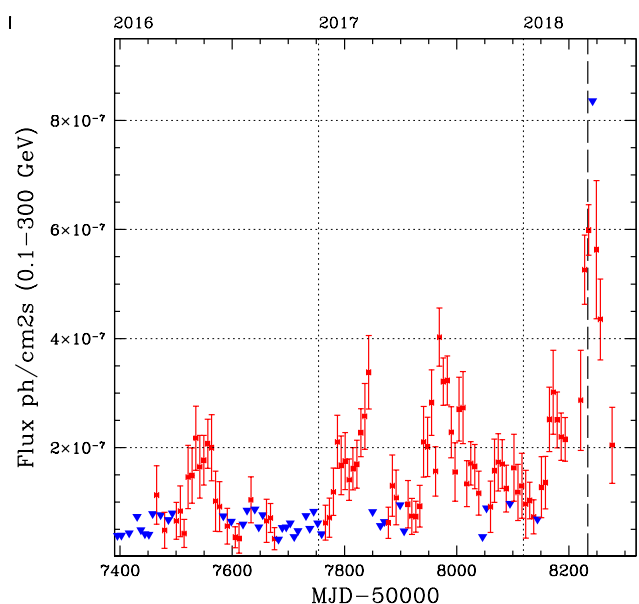}  
\vskip -0.6 cm
\caption{The $\gamma$--ray light curve of TXS 0458--02.
Blue triangles are 95\% upper limits, calculated assuming a power law with photon
spectral index $\Gamma=2$.
The dashed vertical line corresponds to the {\it NuSTAR} observation epoch,
when the source was in a very high $\gamma$--ray state.
} 
\label{0458lat}
\end{figure}

\subsection{TXS 0458--020}

Fig. \ref{0458lat} reports the {\it Fermi}/LAT light curve of the last 3 years,
to show the variability behaviour of this source.
The dashed vertical line indicates the day of the {\it NuSTAR} observation.

\begin{figure} 
 \vskip -0.6  cm
\includegraphics[width=9cm]{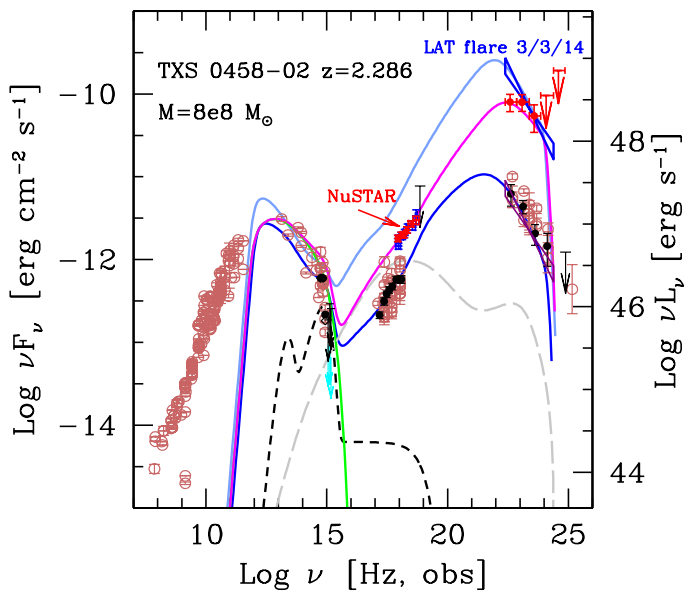}  
\vskip -0.6 cm
\caption{The overall SEDs of TXS 0458--02, showing the
changes in the high energy emission due to its strong variability.
Since, unfortunately, there are no low frequency (mm--optical) data simultaneous
to the varying high energy flux,
the shown models assume a quasi-constant flux at these frequencies.
This illustrates how the model parameters would change in this case.
} 
\label{sed0458}
\end{figure}

Fig. \ref{sed0458} shows the overall SED of the source.
It is characterized by a relatively harder $\gamma$--ray spectrum
with respect to the other two sources, as suggested by the 
nearly simultaneous {\it Fermi}/LAT data (red points).
In this case the flux was high enough to allow the detection
and some spectral determination integrating for one week 
around the {\it NuSTAR} observation. 

Since the synchrotron jet emission hides the accretion disk component,
we cannot fit directly the disk. 
We can derive a (rough) estimate of the accretion disk luminosity
by the  observation of the broad lines, that are seen
in this source even if the continuum is dominated by  the synchrotron
emission.
The CIV broad line has a flux $F_{CIV}=2.6\times 10^{-15}$ erg cm$^{-2}$ s$^{-1}$,
corresponding to a luminosity of $L_{CIV}=1.1\times 10^{44}$ erg s$^{-1}$.
According to the template of Francis et al. (1991) this 
should correspond to a BLR total luminosity of $L_{\rm BLR}=9.7\times 10^{44}$
erg s$^{-1}$ and to a disk luminosity ten times larger: $L_{\rm d} \sim  10^{46}$
erg s$^{-1}$.

For the black hole mass, we must consider that smaller masses, for a given $L_{\rm d}$,
correspond to a disk spectrum peaking at larger frequencies. 
Therefore we can derive an lower limit to the black hole mass requiring
that the disk emission does not over--contribute to the optical--UV flux.
We can have an upper limit to the mass requiring that the disk is emitting
is geometrically thin and optically thick, and therefore has a luminosity
larger than 0.01 $L_{\rm Edd}$.
We have chosen $L_{\rm d}=0.1 L_{\rm Edd}$ for $L_{\rm d} \sim 2\times 10^{46}$ erg s$^{-1}$,
deriving $M=8\times 10^8 M_\odot$.
These values are only indicative, and uncertain by at least a factor 2.

\begin{figure} 
 \vskip -0.6  cm
\includegraphics[width=9cm]{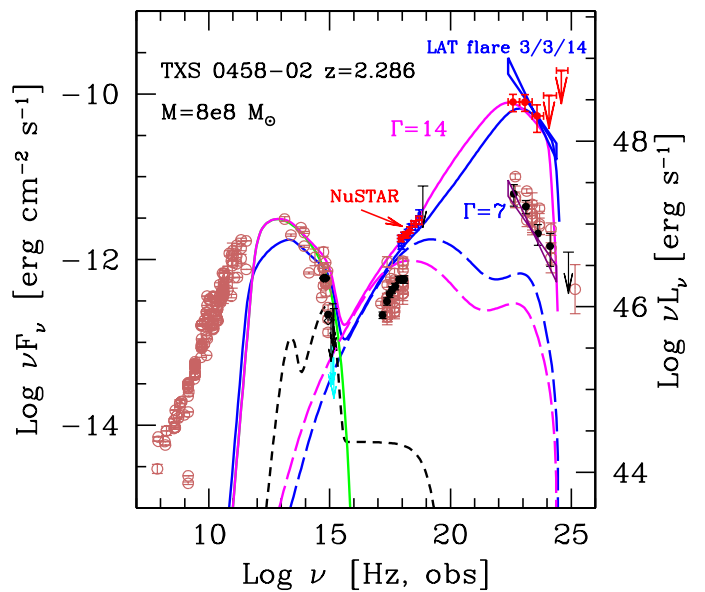}  
\vskip -0.6 cm
\caption{Comparison of the models adopting $\Gamma=14$ and $\Gamma=7$,
as labelled.
Parameters in Table \ref{para} and Table \ref{para2}.
The model with $\Gamma=7$ slightly underestimates the {\it NuSTAR} data.
} 
\label{sed0458g7}
\end{figure}

To explain the observed different states, we have assumed that the archival data
are representative of the quiescent state, while during the {\it NuSTAR} observation
the source was in a high state. 
In March 2014 there was a {\it Fermi}/LAT flare almost brighter than in 2018,
but unfortunately with no other observations at other frequencies.
We show a possible fit for this flare, but only to illustrate the change of the parameters
if the source would ever resemble the proposed theoretical SED.

As usual, we look for solution involving the smallest change of the minimum number of parameters
to explain the observed variability.
For the {\it ``NuSTAR} state" the power injected in relativistic electrons if 4 times 
larger than in the quiescent state, but the magnetic field is $\sim$2.5 times smaller.
The slopes of the injected electron distribution are slightly harder and
the total jet power, in the {\it ``NuSTAR"} state, is twice as much than in quiescence.
The ``high" state would require more power in the injected electrons (more than 10 times than
in quiescence), a smaller still magnetic field, and the total jet power would be $\sim$13 times
larger.
All these estimates are calculated assuming that the synchrotron part of the spectrum 
is well represented by the quiescent state, in turn shown by the archival data.
This source was studied also in Ghisellini et al. 2011, where simultaneous {\it Swift}
(UVOT and XRT) and {\it Fermi}/LAT observations are reported. 
They correspond to the black symbols in Fig. \ref{sed0458}.

Recently, Lister et al. (2016) have measured the apparent speed of a superluminal knot in
this source, deriving an apparent speed $\beta_{\rm app}\sim 6$.
Although this is a lower limit to the value of the bulk Lorentz factor, and therefore
consistent with the values used in Fig. \ref{sed0458}, it is interesting to compare these
models with the one using a smaller value of $\Gamma$.
This is done in Fig. \ref{sed0458g7}, that compares the models with $\Gamma=14$ and $\Gamma=7$,
as labelled.
The latter slightly underestimate the {\it NuSTAR} data, but can reproduce well the rest
of the SED.
The parameters listed in Table \ref{para} and Table \ref{para2} indicate (for the $\Gamma=7$ case)
that the jet power  and the magnetic field are slightly smaller, and the electron energies are larger.
Overall, we have that the parameters are not vastly different.

\section{Discussion}

\subsection{Comparison with other $z>2$ {\it NuSTAR} blazars}

\begin{figure} 
\vskip -0.6 cm
\includegraphics[width=9.5cm]{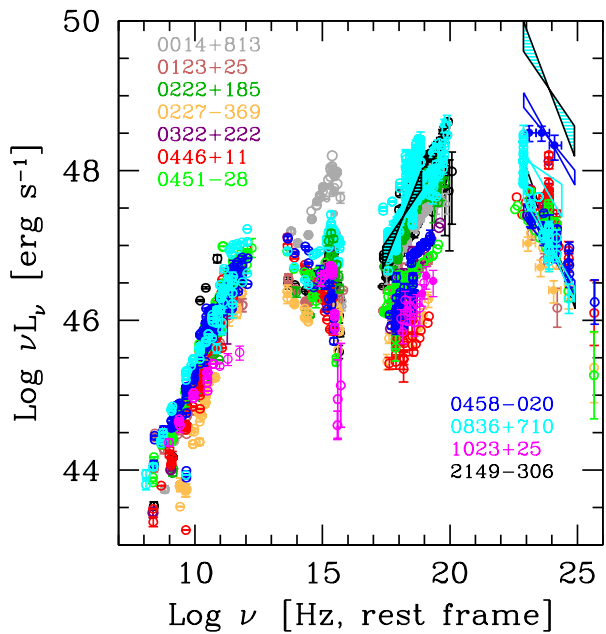}  
\vskip -0.6 cm
\caption{The SED of all 11 blazars at $z>2$ observed so far by {\it NuSTAR}.
It can be noted that 1) the synchrotron hump is remarkably similar;
2) for several sources the accretion disk sticks out in the optical--UV band;
3 0014+813 has an exceptionally powerful accretion disk;
4) the X and $\gamma$--ray emission is more dispersed and variable.
} 
\label{allsed}
\end{figure}

Tab. \ref{all} reports the list of all blazars at $z>2$ observed by {\it NuSTAR}.
They are 11 sources. 
The table reports their redshift and the reference to the papers discussing
the {\it NuSTAR} X--ray data. 
They all are FSRQs, and their SEDs are shown in Fig. \ref{allsed},
in the $\nu L_\nu$ vs $\nu$ (rest frame) representation.
In this way we can compare the rest frame SED of the sources.
Most of the data comes from archives (mostly ASI/SSDC)
and the figure shows how similar the sources are
in the radio--mm band, while they become different (and varying with
a very large amplitude) at greater frequencies.
Note the source S5 0014+813, the most luminous in the optical--UV,
due to its extraordinary luminous accretion disk (Ghisellini et al. 2009), and
S5 0836+710, the most luminous in X--rays and in $\gamma$--rays,
where it reached a luminosity of $\sim 10^{50}$ erg s$^{-1}$
during a flare observed on August 2nd, 2015 (Ciprini 2015).

\begin{table} 
\centering
\footnotesize
\begin{tabular}{lllll lllll ll}
\hline
\hline
Name &$z$     &Ref \\
\hline   
S5 0014+81    &3.366  &S16, B18 \\  
PKS 0123+25   &2.358  &This paper \\ 
B0222+185     &2.690  &S16, B18 \\
PKS 0227--369 &2.115  &This paper\\ 
TXS 0322+222  &2.066  &M17    \\       
PKS 0446+11   &2.15   &M17    \\  
PKS 0451--28  &2.564  &M17 \\      
TXS 0458--020 &2.291  &This paper \\ 
S5 0836+710   &2.172  &T15, P15, B18 \\
B2 1023+25    &5.3    &S13 \\
PKS 2149--306 &2.345  &T15, D16, B18\\
\hline
\hline 
\end{tabular}
\caption{The entire sample of $z>2$ blazars observed by {\it NuSTAR}.
References: 
S16: Sbarrato et al. 2106;
B18: Bhatta et al. 2018;
S16: Sbarrato et al. 2016;
M17: Marcotulli et al., 2017;
T15: Tagliaferri et al., 2015,
P15: Paliya et al., 2015;
S13: Sbarrato et al. 2013;
D16: D'Ammando \& Orienti, 2016.
}
\label{all}
\end{table}

\begin{figure} 
\vskip -0.4 cm
\includegraphics[width=9.5cm]{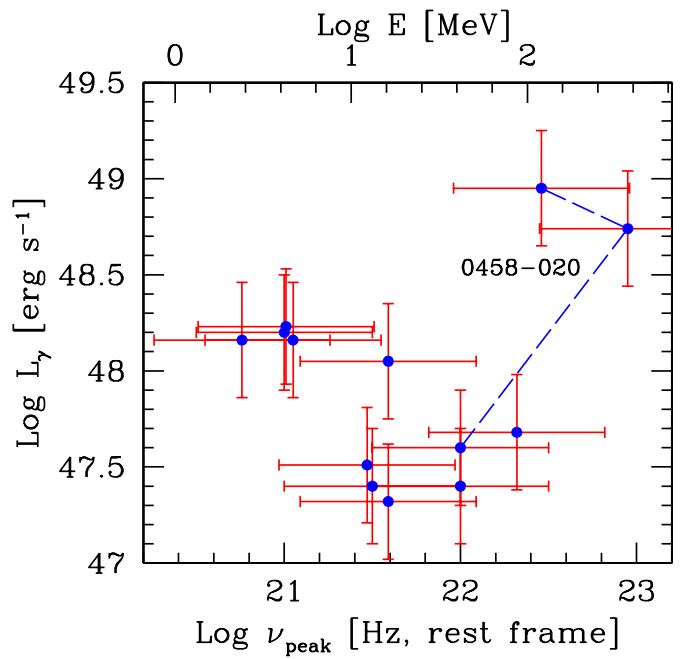}  
\vskip -0.6 cm
\caption{
Peak luminosity of the high energy component as a function
of its peak frequency. 
The dashed line connects three different states of TXS 0458--020.
Error bars correspond to a factor 3 uncertainties in $\nu_{peak}$
and a factor 2 in $L_\gamma$.
There is a (weak) trend of smaller luminosities for larger peak frequencies,
with the exception of TXS 0458--020 when in the high state.
} 
\label{vplg}
\end{figure}

The reason of the smaller dispersion of data points in the radio
with respect to the other wavelengths is probably due to the
lower amplitude variability in the radio band. 
Another reason for having less dispersion in the radio--mm band
is that the Doppler amplification of the synchrotron flux
scales as $F(\nu) \propto \delta^{3+\alpha}\sim \delta^3$
(for flat spectral indices $\alpha=0$),
while the amplification factor for the inverse Compton process, with
photons produced externally to the jet, scales as $F(\nu) \propto \delta^{4+2\alpha}\sim \delta^5$
(for X-ray spectral indices $\alpha\sim 0.5$)
as pointed out by Dermer (1995) and illustrated in Fig. 5 of Ghisellini (2015).

Note that all sources show no signs of changing slope at the lowest
radio frequencies, an indication that the jet emission is extremely strong
and hides any contribution of the extended radio structure, that should
have a steep (i.e. increasing at lower frequencies) spectrum.
On the other hand, for almost all sources we do see the contribution of the
accretion disk in the optical--UV.
The accompanying X--ray coronal emission is absent in these sources, 
completely overwhelmed by the beamed X--rays from the jet.
As a consequence for no source there is any sign of the presence of the
iron fluorescence line at 6.4 keV (rest frame).

The hardness of the X--ray spectrum coupled with the steepness 
of the $\gamma$--ray one indicates a spectral peak around $\sim$10 MeV.
We can try to be more precise by extrapolating the X--ray and $\gamma$--ray 
spectra of each source and find out the matching frequency.
The result is shown in Fig. \ref{vplg}: the $\gamma$--ray luminosity $L_\gamma$
is plotted vs the peak frequency.
For $L_\gamma$ we have chosen an average state, not the extreme 
flaring state.
Bear in mind that this result can be affected by systematic errors,
since the spectral shape around the peak is likely to be curved
and not accurately described by a broken power law.
The figure in any case suggests a trend (smaller $\nu_{peak}$ for larger $L_\gamma$)
and an outlier (TXS 0458--020 in the high state).

\begin{figure} 
\vskip -0.6 cm
\includegraphics[width=9.5cm]{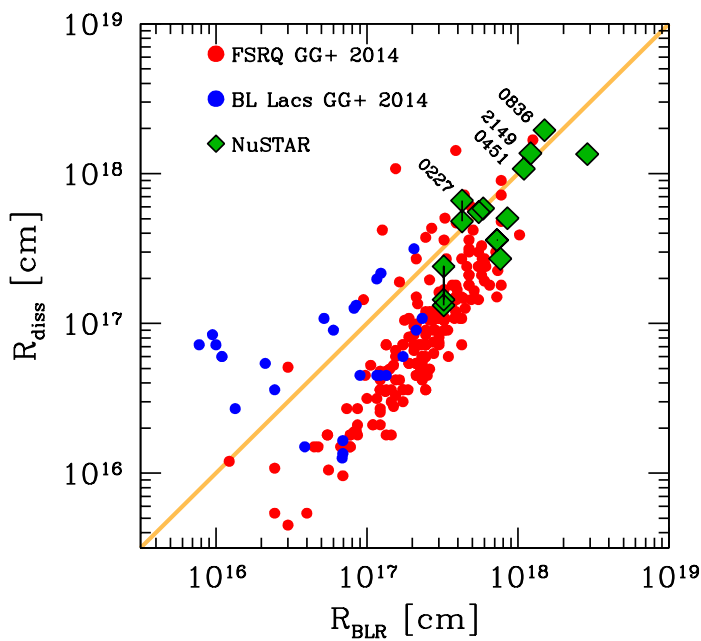}  
\vskip -0.6 cm
\caption{The distance $R_{\rm diss}$ at which most of the luminosity is produces
as a function of the size of the broad line region, $R_{\rm BLR}$.
Blue (``BL Lacs", but having broad emission lines, see text) and red (FSRQ) 
from Ghisellini et al. (2014).
Green diamonds are our {\it NuSTAR} blazars. 
Different states of the same source are connected by a segment.
For about 12\% of all sources the dissipation region is located beyond 
the BLR ($R_{\rm diss}>R_{\rm BLR}$).
} 
\label{rdiss}
\end{figure}
\subsection{Seed photons from the BLR or the torus?}

The peak frequency $\nu_{\rm C}$ of the high energy hump of blazars depends on 
the frequency of the seed photons, the energy of the relevant electrons
contributing to the peak, the bulk Lorentz factor $\Gamma$ 
and the beaming factor $\delta$.
For our sources, that are all very powerful, we can assume that $\delta\sim\Gamma$,
that implies that the viewing angle $\theta_{\rm v}\sim 1/\Gamma$.
If the emitting region is inside the broad line region (i.e. $R_{\rm diss}< R_{\rm BLR}$
the most important seed photons are the Ly$\alpha$ ones.
Therefore we expect
\begin{equation}
\nu_{\rm C}  \, =\, {4\over 3}\gamma^2_{\rm peak} \nu_{\rm Ly\alpha} {\Gamma^2 \over 1+z};
\qquad R_{\rm diss}< R_{\rm BLR}
\label{vc1}
\end{equation}
If $R_{\rm BLR} < R_{\rm diss} < R_{\rm torus}$ the most important seed photons 
are the ones produced by the torus.
These have a frequency related to the torus temperature, that has to be
smaller than $\sim$2000 K to avoid sublimation.
\begin{equation}
\nu_{\rm C}  \, =\, {4\over 3}\gamma^2_{\rm peak} \nu_{\rm torus} {\Gamma^2 \over 1+z};
\qquad R_{\rm BLR} < R_{\rm diss} < R_{\rm torus}
\label{vc1}
\end{equation}
%
\begin{figure} 
\vskip -0.6 cm
\includegraphics[width=9.5cm]{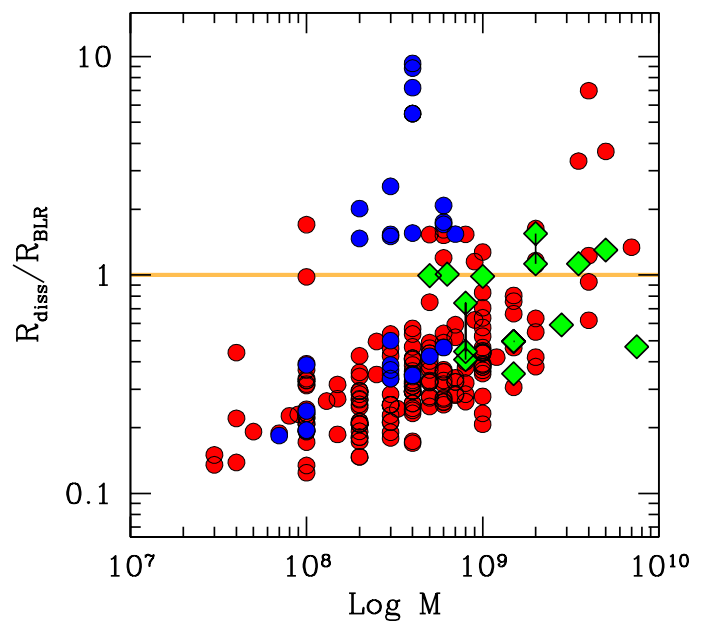}  
\vskip -0.6 cm
\caption{The ratio $R_{\rm diss}/R_{\rm BLR}$ as a function of the
black hole mass.
Blue (``BL Lacs") and red (FSRQ) 
from Ghisellini et al. (2014). Green diamonds are our {\it NuSTAR} blazars. 
Different states of the same source are connected by a segment.
} 
\label{rdiss_m}
\end{figure}
%
The ratio of the two $\nu_{\rm C}$ frequencies is $\sim 40 \, (10^3\, {\rm K}/T_{\rm torus})$.
If the emitting region is at a distance greater, but close to $R_{\rm BLR}$,
both types of seed photons are important, and we have an intermediate peak frequency
as long as $\gamma_{\rm peak}$ is the same. 
In general, one would expect that the radiative cooling time 
is affected by the nature of the seed photons: inside the BLR the
BLR radiation energy density is larger than the one produced by the torus.
Cooling is more severe, and this could favour smaller $\gamma_{\rm peak}$.
This compensates the larger seed photon energy. 
On the other hand, we calculate the particle distribution at the end
of the injection, that lasts for a time $R/c$. 
We also assume that the jet is conical, and therefore $R\propto R_{\rm diss}$:
if the emitting region is beyond $R_{\rm BLR}$, it is larger than
if it is inside.
This means that emission (and cooling) operates for a longer time,
and this has the effect to decrease $\gamma_{\rm peak}$.
So, it is not obvious that sources dissipating beyond $R_{\rm BLR}$
should be ``bluer" than the others.
In any case, we have tried to see for how many blazars 
studied previously by our group we require $R_{\rm diss}>R_{\rm BLR}$.

\begin{figure} 
\vskip -0.6 cm
\includegraphics[width=9.5cm]{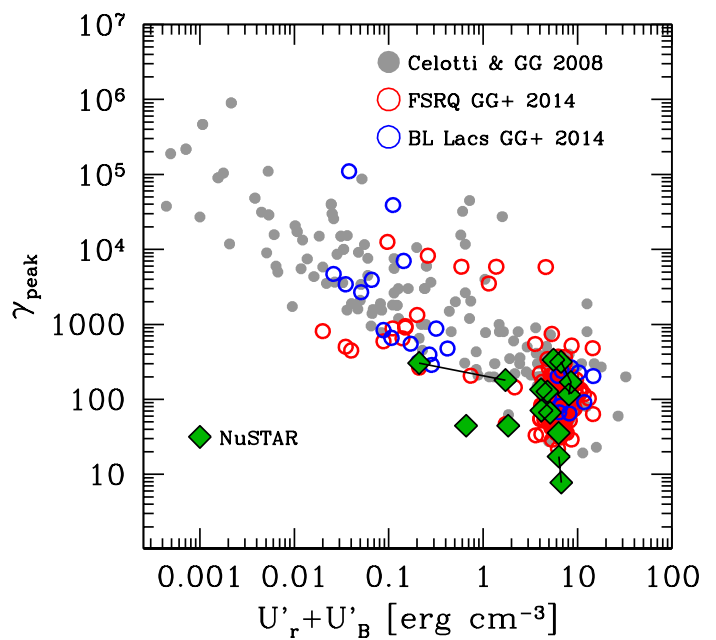}  
\vskip -0.6 cm
\caption{
Random Lorentz factor of the electrons emitting at the synchrotron
and IC peaks vs the radiation+magnetic energy density as measured in the
comoving frame.
Grey filled circles: sources studied in Celotti et al. (2008); 
empty red circles and blue circles: FSRQs and ``BL Lacs"
from Ghisellini et al. (2014);
green diamonds: the sample of $z>2$ blazars observed by {\it NuSTAR}.
Segments connect different states of the same source.
} 
\label{gpeak}
\end{figure}
Fig. \ref{rdiss} shows $R_{\rm diss}$ as a function of $R_{\rm BLR}$ 
for the sample of blazars studied in Ghisellini et al. (2014)
and for the high redshift {\it NuSTAR} FSRQs studied here.
The figure shows that there is a small ($\sim$12\%) fraction of sources
having $R_{\rm diss}\gsim R_{\rm BLR}$
and that there is an overall trend for $R_{\rm diss}$ increasing more
than linearly with $R_{\rm BLR}$.
The {\it NuSTAR} blazars requires the largest $R_{\rm diss}$ and $R_{\rm BLR}$
and nearly half of them dissipate beyond $R_{\rm BLR}$.
We can also wonder if the $R_{\rm diss}/ R_{\rm BLR}$ ratio
is a function of the black hole mass. 
We do expect some dependence, because $R_{\rm BLR}$ depends on the black hole mass
only through $L_{\rm d}$ (and we do expect more luminous disk for larger black hole masses),
while $R_{\rm diss}$ should scale linearly with the mass if dissipation occurs
at the same distance measured in units of the \sch\ radius.
Therefore we expect a dependence (albeit weak) for larger ratios $R_{\rm diss}/R_{\rm BLR}$ 
for larger masses.
Fig. \ref{rdiss_m} shows this weak trend.

\begin{figure} 
\vskip -0.6 cm
\includegraphics[width=9.5cm]{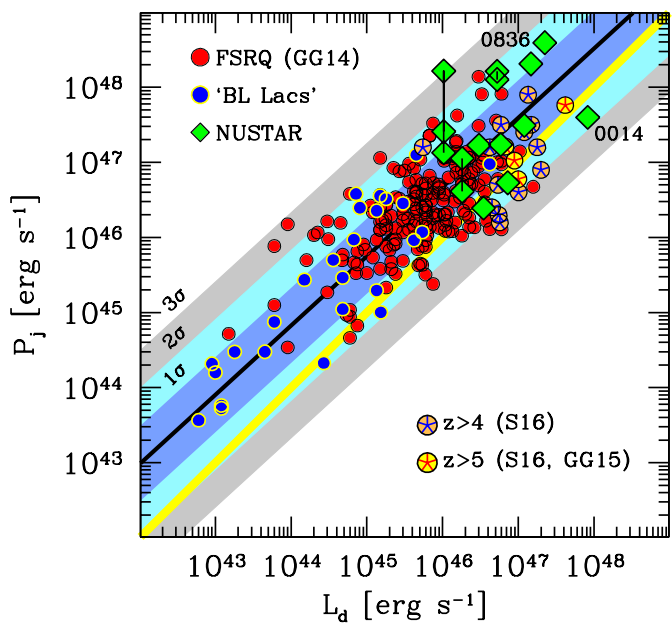}  
\vskip -0.6 cm
\caption{Jet power as a function of disk luminosity of FSRQs (red)
and ``BL Lacs" (blue) considered in Ghisellini et al. (2014)
compared with the {\it NuSTAR} blazars considered here.
We also show the blazars with $z>4$ and $z>5$, considered
in Sbarrato et al. (2016) and in Ghisellini et al. (2015).
Segments connect different states of the same source. 
The {\it NuSTAR} blazars are among the most powerful, both in terms
of their disk luminosity and jet power,
with PKS 0836+710 having the most powerful jet, 
and S5 0014+81 having the most powerful accretion disk.
Note that the BL Lacs shown here were the only BL Lacs
observed by Shaw et al. (2013) with broad emission lines.
They must be considered the low disk luminosity tail of FSRQ.
} 
\label{pjet}
\end{figure}

\subsection{$\gamma_{\rm peak}$--$U^\prime$ relation}

We now consider the relation between the electron random Lorentz factor $\gamma_{\rm peak}$ of
the electrons emitting at the peaks of the SED (both synchrotron and IC)
and the magnetic plus radiation energy density in the comoving frame of the emitting region.
This is shown in Fig. \ref{gpeak} that compares our high--$z$ {\it NuSTAR}
blazars with the samples of blazars studied by Celotti \& Ghisellini (2008)
and Ghisellini et al. (2014).
If considered altogether, there is a clear trend of decreasing $\gamma_{\rm peak}$
for increasing energy density.
On the other hand, the number of {\it NuSTAR} blazars is too small to
derive any conclusions: they are, as all the other powerful FSRQs, at
the extreme of the distribution.

\subsection{Jet power and disk luminosity}

Finally, in Fig. \ref{pjet}, we consider the jet power as a function of the disk luminosity.
The blue circles are labeled ``BL Lacs", as it was done in 
(Ghisellini et al. 2014).
They come from the sample of Shaw et al. (2013), containing 475 sources.
Of these, Ghisellini et al. (2014) selected the few (26) objects with broad emission
lines. 
Therefore these ``BL Lacs" should be considered as the low disk luminosity 
tail of FSRQs.
The relation between $P_{\rm jet}$ and $L_{\rm d}$ remains significant even after accounting 
for the common dependence upon redshift, with a probability $P < 10^{-8}$ to be random
(Ghisellini et al. 2014).
This figure clearly shows that the {\it NuSTAR} blazars studied in this paper
are the most powerful.
This remains true even if we consider the {\it lower limit} to the jet power
given by $P_{\rm r}$, that is almost model independent.
PKS 0836+710 has the most powerful jet, 
and S5 0014+81 has the most powerful accretion disk.
They extend the almost linear correlation between the two quantities found in 
Ghisellini et al. (2014), and confirm that active blazars have jets often
more powerful than their accretion disks.

\section{Conclusions}

\begin{itemize}

\item Selection in the hard X--rays allows to find the most powerful blazar
jets and the most luminous accretion disk.

\item PKS 0227--369 and TXS 0458--020 show a significant variability
in hard X--rays with respect to previous observations.
This variability can be explained mainly by a change of power of the injected electrons
and in part by a change of the magnetic field.

\item All the high--$z$ {\it NuSTAR} blazars observed so far belong to the class
of very powerful FSRQs and have large black hole masses and accretion disks emitting well
above the 0.01 $L_{Edd}$ rate.


\item
The high--$z$ {\it NuSTAR} blazars extend and confirm the relation 
between jet power and accretion disk luminosity.

\end{itemize}

%
%
%

\section*{Acknowledgements}
We acknowledge the ASI--{\it NuSTAR} grant ASI 1.05.04.95 and the grant
ASI-INAF n. 2017--14--H.0 for funding.




\end{document}